\documentclass[aps,prl,twocolumn,superscriptaddress,showpacs,english]{revtex4-1}

\usepackage[T1]{fontenc}
\usepackage[latin9]{inputenc}
\usepackage{babel}
\usepackage{amsmath}
\usepackage{amssymb}
\usepackage{wasysym}
\usepackage{graphicx}
\usepackage{xcolor}
\usepackage{graphicx}

\usepackage[linktocpage=true,
  colorlinks=true, 
  pdfborder={0 0 0},
  linkcolor=blue,
  citecolor=red,
  filecolor=yellow,
  urlcolor=blue,
  bookmarks,
  pdfauthor={},
]{hyperref}

\newcommand{\Basel}{Department of Physics, Universit\"at Basel, Klingelbergstr. 82, 4056 Basel, Switzerland}
\newcommand{\Halle}{Max-Planck Institut f\"ur Microstrukture Physics, Weinberg 2, 06120 Halle, Germany}
\newcommand{\Evanston}{Department of Materials Science and Engineering, Northwestern University, Evanston, IL 60208, United States}
\newcommand{\Graz}{Institute of Theoretical and Computational Physics, Graz University of Technology, NAWI Graz, 8010 Graz, Austria}
\newcommand{\Aquila}{Dipartimento di Fisica Universit\`{a} degli Studi di L'Aquila and SPIN-CNR, I-67100 L'Aquila, Italy}

\newcommand{\tc}{T$_{\textmd C }$}

\newcommand{\omlog}{$\omega_{\textmd log}$}

\begin{document}

\title{Superconductivity in metastable phases of phosphorus-hydride compounds under high pressure}

\author{Jos\'e A. Flores-Livas} \affiliation{\Basel} 
\author{Maximilian Amsler}      \affiliation{\Evanston}
\author{Christoph Heil}         \affiliation{\Graz}
\author{Antonio Sanna}          \affiliation{\Halle}
\author{Lilia Boeri}            \affiliation{\Graz} 
\author{Gianni Profeta}         \affiliation{\Aquila}
\author{Chris Wolverton}        \affiliation{\Evanston}
\author{Stefan Goedecker}       \affiliation{\Basel}
\author{E.~K.~U. Gross}         \affiliation{\Halle}

\date{\today}

\begin{abstract}
Hydrogen-rich compounds have been extensively studied both theoretically and experimentally in the quest for novel high-temperature superconductors. 
Reports on sulfur-hydride attaining metallicity under pressure and exhibiting superconductivity at temperatures as high as 200\,K 
have spurred an intense search for room-temperature superconductors in hydride materials. 
Recently, compressed phosphine was reported to metallize at pressures above 45\,GPa, reaching a superconducting transition 
temperature (\tc) of 100\,K at 200\,GPa. However, neither the exact composition nor the crystal structure of the superconducting phase have been conclusively determined. 
In this work the phase diagram of PH$_n$ ($n=1,2,3,4,5,6$) was extensively explored by means of {\it ab initio} crystal structure 
predictions using the Minima Hopping Method (MHM). 
The results do not support the existence of thermodynamically stable PH$_n$ compounds, which exhibit a tendency for elemental 
decomposition at high pressure even when vibrational contributions to the free energies are taken into account. 
Although the lowest energy phases of PH$_{1,2,3}$ display \tc's comparable to experiments, it remains uncertain if the measured 
values of \tc \ can be fully attributed to a phase-pure compound of PH$_n$.
\end{abstract}

\pacs{~}
\maketitle

In December 2014, {\em Drozdov et al.} reported a superconducting critical temperature (\tc) of 203 K in an ultra-dense phase of sulfur-hydride 
(SH$_3$)~\cite{DrozdovEremets_Nature2015},  identified by \textit{ab initio} crystal structure searches~\cite{Duan_SciRep2014},
breaking the record-\tc \ previously held by the cuprates. 
Subsequent experimental and theoretical studies have confirmed that superconductivity is of conventional nature and occurs 
in the predicted bcc phase~\cite{DrozdovEremets_Nature2015,Drozdov_XRAY}, 
demonstrating the potential of {\em ab-initio} crystal structure search methods to identify new superconductors.
Several studies have meanwhile appeared in literature, discussing different aspects underlying the exceptional \tc, such as the
role of bonding, Coulomb screening, phonon anharmonicty, etc.~\cite{SH_PRB-Mazin-2015,Errea_anhaPRL2015,FloresSanna_H3Se_Arxiv2015,PRB_Duan2015,PRB_akashi_2015_HS,Heil-Boeri_PRB2015,quan_impact_2015,ortenzi_TB_2015}.

High-\tc \ superconductivity based on a conventional electron-phonon ($ep$) coupling mechanism has been suggested by Ashcroft almost fifty years ago. 
He originally proposed that this could be achieved if hydrogen was sufficiently compressed, 
a prediction that has not yet been verified due to the required extreme pressures~\cite{Ashcroft_PRL1968,Cudazzo_PRL2008,Cudazzo_1_PRB2010,Cudazzo_2_PRB2010,Szcz_superconducting_2009,mcmahon_high_2011}. 
More recently, he suggested that the chemical pre-compression of hydrogen in hydrogen-rich compounds could be an effective route 
to reach metallization and high-\tc \ superconductivity at experimentally accessible pressures~\cite{Ashcroft_PRL2004}, 
stimulating an intense activity of \textit{ab initio} searches and predictions for high-\tc \ superconducting 
hydrides~\cite{tse_novel_2007,Chen_PNAS2008,Kim_PNAS2008,FengAsHoffman_Nature2008,Wang_PNAS2009,Yao_PNAS2010,gao_high-pressure_2010,Kim_PNAS2010,Li_PNAS2010,Zhou_PRB2012,Hooper_JPC-2014,Duan_SciRep2014,Maramatsu-Hemley_2015,Disilane_JAFL}. 
Until 2014, however, all high-pressure phases which have been synthesized experimentally exhibited rather 
low \tc~\cite{Eremets_Science2008,degtyareva_formation_2009,Hanfland_PRL2011,PtH_PRL2011}.

Eventually, the discovery of SH$_3$ showed that high-\tc \ in hydrogen-rich solids can indeed be achieved. 
Thus, all stable hydrogen-containing molecules which can be placed into the compressing chamber of a diamond anvil cell are potential candidates for high-\tc \ superconductors, 
as long as they remain stable against decomposition, amorphization, and the possible formation of metal-hydrides 
during the measurement of \tc~\cite{Eremets_Science2008,degtyareva_formation_2009,Hanfland_PRL2011,PtH_PRL2011}.

In fact, less than one year after the discovery of superconductivity in SH$_3$, 
Drozdov~\textit{et al.} have very recently reported high-\tc \ superconductivity 
in a second hydrogen-rich compound at extreme pressures: resistivity measurements on phosphine (PH$_3$) show that the samples, 
which are semiconducting at ambient pressure, metallize above 40\,GPa and become superconducting at around 80\,GPa, 
exhibiting a maximum \tc \ of 100\,K at about 200\,GPa~\cite{Drozdov_ph3_arxiv2015}. 
Neither the exact composition of the superconducting phase and its crystal structure, nor the mechanism responsible for 
the high-\tc \ have been conclusively determined at this point. 
Analogies with superconducting SH$_3$, which was obtained from SH$_2$ precursor, 
suggest that the superconductivity in the P-H system is of conventional nature, but that the composition of the superconducting 
phase might be different from the original PH$_3$ stoichiometry. 

To shed light on this matter we used \textit{ab initio} techniques to map out the high-pressure phase diagram 
of the P-H binary system by exploring the compositional and configurational space of PH$_n$  with a sophisticated structure prediction method, 
and estimated the superconducting properties of the most promising phases. 
We found that all high-pressure binary phases of P and H are metastable with respect to elemental decomposition in the pressure range 
100-300 GPa. However, the critical temperatures of the three phases closest to the convex hull (PH, PH$_2$ and PH$_3$) reproduce 
to a good approximation the experimental \tc~values. 
Possible ways to reconcile our results with experiments are discussed at the end of this manuscript.

\begin{figure}[t]
\includegraphics[width=1.0\columnwidth,angle=0]{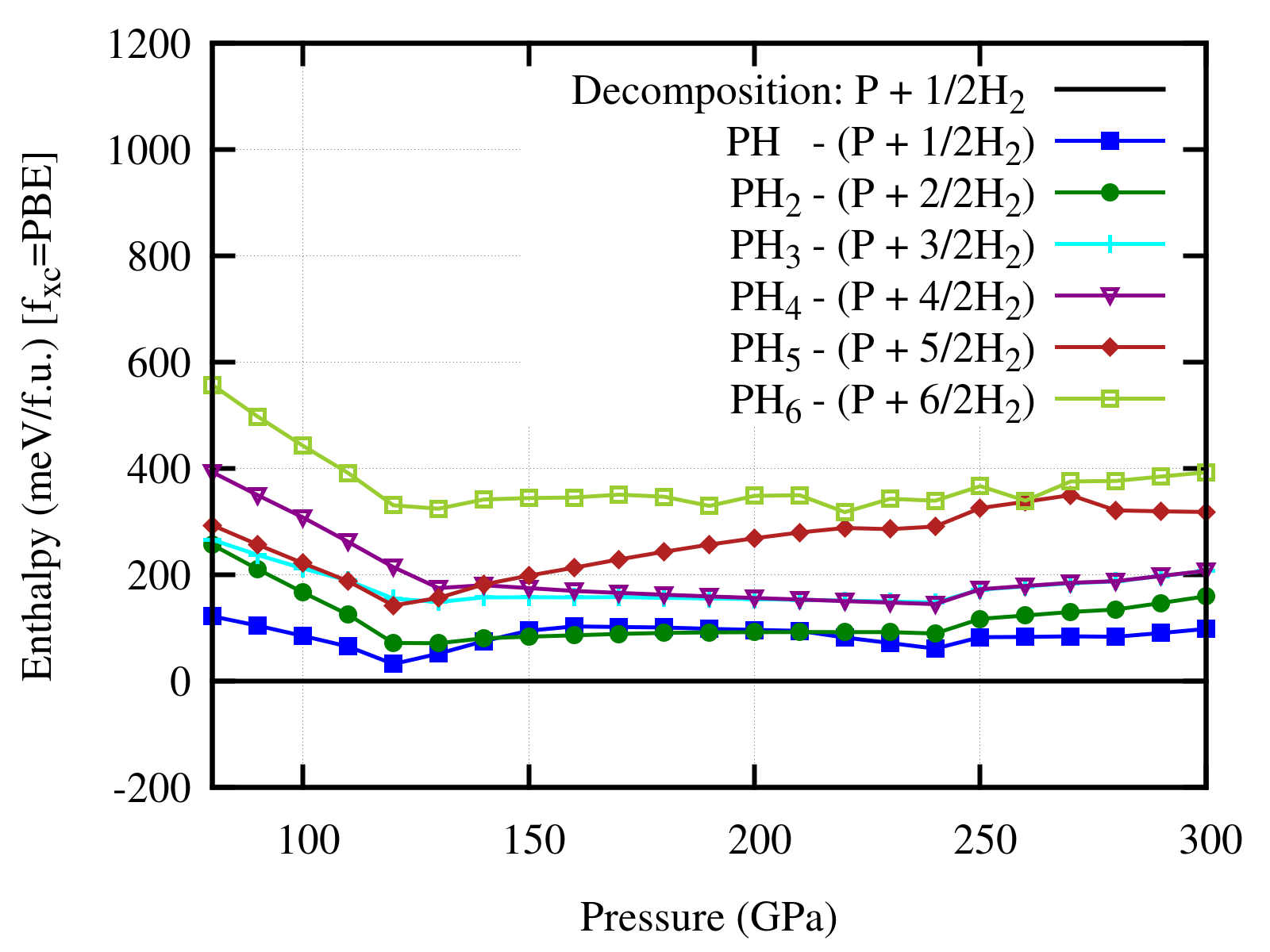}
\caption{(Color online) Calculated enthalpies for PH$_n$ ($n=1,2,3,4,5,6$) for the low-lying structures found in this work. 
 Values are given with respect to the elemental decomposition (P + 1/2H$_2$). 
 The reference structures for hydrogen are $P6_3m$ (0-120\,GPa) and $C2/c$ (120-300\,GPa) from Ref.~\cite{pickard_structure_2007}. 
 The reference phases for phosphorus   are: $Pm\bar{3}m$ (0-10\,GPa), $R3m$ (20-110\,GPa), $P6mm$ (120-240\,GPa), 
 $Im\bar{3}m$ (250-270\,GPa) and $I\bar{4}3d$ (280-300\,GPa). 
 The change in slope at around 120\,GPa is due to the phase transition in elemental hydrogen.}
\label{fig:enthalpy}
\end{figure}

To sample the enthalpy landscape we employed the Minima Hopping method (MHM)~\cite{Goedecker_2004,Amsler_2010}, 
which has been successfully used for global geometry optimization~\footnote{The MHM was designed to thoroughly scan the 
low-lying enthalpy landscape of any compound and identify 
stable phases by performing consecutive short molecular dynamics escape steps followed by local geometry relaxations. 
The enthalpy surface is mapped out efficiently by aligning the initial molecular dynamics velocities approximately along 
soft mode directions~\cite{roy_2009,sicher_efficient_2011}, thus exploiting the Bell-Evans-Polanyi~\cite{jensen_introduction_2011} 
principle to steer the search towards low energy structures.} in a large variety of applications~\cite{MA_JAFL,LiAlH_Maxmotif,BJAFL_PRB2012}, 
including superconducting materials at high pressure~\cite{Disilane_JAFL,FloresSanna_H3Se_Arxiv2015}. 
The enthalpies as a function of pressure for the ground state structures~\footnote{We explored phases with 1, 2, 3 and 4 formula units of PH$_n$ for stoichiometries $n=1,2,3,4,5,6$ at selected pressures in 
the range 100-300 GPa. The relaxations to local minima were performed by the fast inertia relaxation engine~\cite{FIRE_2006} 
by taking into account both atomic and cell degrees of freedom. Energy, atomic forces and stresses were evaluated at the density functional theory (DFT) 
level with the Perdew-Burke-Erzernhof (PBE)~\cite{PBE96} parametrization to the exchange-correlation functional. 
A plane wave basis-set with a high cutoff energy of 1000\,eV was used to expand the wave-function together with 
the projector augmented wave (PAW) method as implemented in the Vienna Ab Initio Simulation Package~{\sc vasp}~\cite{VASP_Kresse}. 
Geometry relaxations were performed with tight convergence criteria such that the forces on the atoms were less than 2~meV/\AA~and 
the stresses were less than 0.1~eV/\AA$^3$.} at each composition are shown in Fig.~\ref{fig:enthalpy}  with respect to elemental decomposition. 
The ground state structures were determined by the MHM at 100, 150, 200 and 300\,GPa, and further relaxed at intermediate pressures 
to obtain a smooth phase diagram. An additional search was carried out at zero pressure. 
The various phases of pure H and P for the convex hull construction were obtained from a structural search, 
and for hydrogen they coincide with the ones reported by Needs~\textit{et al.}~\cite{pickard_structure_2007}: $P6_3m$ (0-120\,GPa) and $C2/c$ (120-300\,GPa).  
The complex phase diagram of elemental phosphorus~\cite{Sugimoto_Psuperlattice-PRB-2012,marques_origin_2008},
with at least six structural transition between 0 and 300~GPa, is closely reproduced by our calculations - see supplemental Material.
At zero pressure, molecular phosphine (PH$_3$), as well as the crystalline phases of PH, PH$_2$ and PH$_3$ 
identified by our MHM runs, are all stable with respect to elemental decomposition (see supplemental materials). 
Moreover, we found two compositions (PH and PH3) forming the convex hull with enthalpies of formation between -50 and -60 meV/atom.
In the crystalline phases, the molecules retain their geometry and are held together by van-der-Waals interactions; 
these molecular crystals are semiconducting with gaps of 0.9 (PH), 2.1 (PH$_2$) and 3.7\,eV (PH$_3$, respectively. 
Given the complexity of the phase diagram of elemental phosphorus, 
it is likely that binary phases considered here will undergo several phase transitions between ambient 
pressure and 100\,GPa, but analyzing these transitions is well beyond the scope of this work. 
Assuming that the PH$_3$ zero-pressure structure is representative
at higher pressures, we estimate a metallization pressure of $\sim$\,35\,GPa, which is in accordance 
with experiments -- see supplemental materials for details. 

In the following, we will focus on high pressures (80-200\,GPa), which are relevant for superconductivity.  
The formation enthalpies of all the identified structures are positive and in a range of 
30-200 meV/f.u. for hydrides with low hydrogen contents (PH, PH$_2$ and PH$_3$), and larger for higher H content up to PH$_6$. 

\begin{figure}[t]
\includegraphics[width=1.0\columnwidth,angle=0]{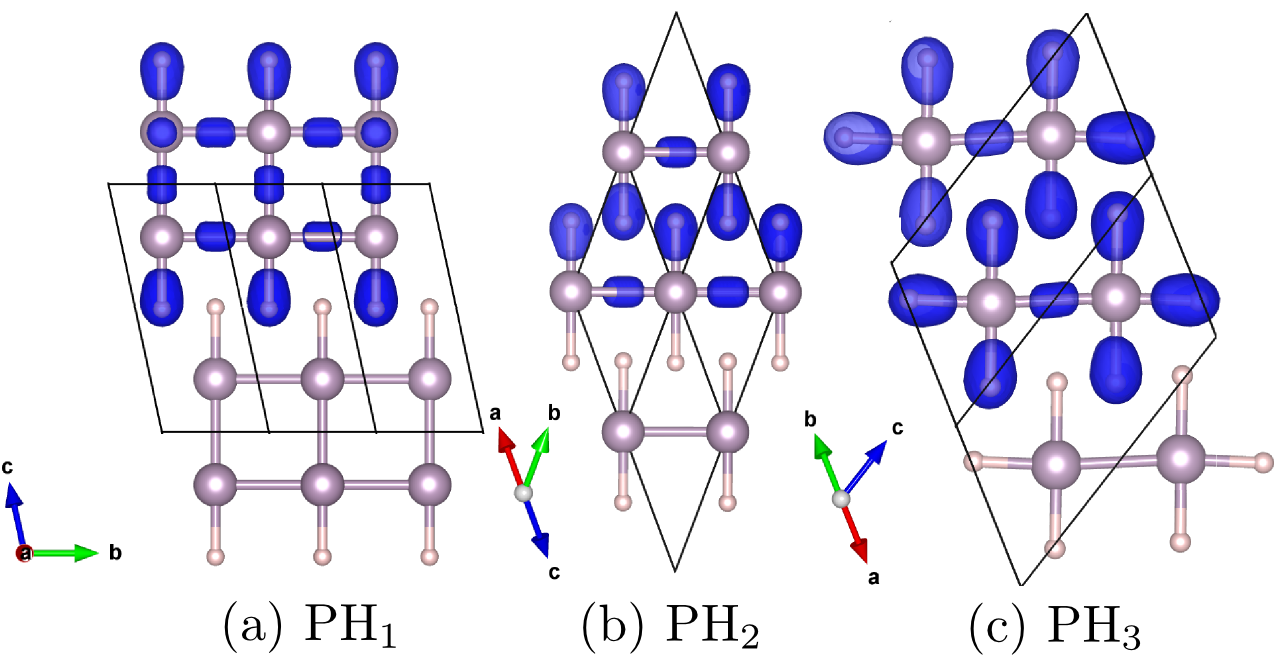}
\caption{(Color online) Low-lying enthalpy structures found for different compositions under pressure at 120\,GPa. 
The large and small spheres denote the P and H atoms, respectively. 
The ELF at a fixed value of 0.8 is shown in the upper part of each structure.}
\label{fig:structures}
\end{figure}

The ground state structures at 120\,GPa of pressure for PH ($I4/mmm$), PH$_2$ ($I4/mmm$) and PH$_3$ ($C2/m$) are shown in Fig.~\ref{fig:structures}. 
The bond lengths are $\sim$\,1.4\,\r{A} for the P-H bonds and $\sim$\,2.1\,\r{A} for the P-P bonds in all three structures. 
Figure~\ref{fig:structures} clearly shows the dominating tendency of P atoms to form polymers with hydrogen saturating the dangling bonds, 
resulting in 1D chains or 2D layers. This picture is confirmed when looking at the iso-surfaces of the electron localization function (ELF), 
clearly showing that electrons are localized to form P-P and P-H bonds. 
No or only weak interactions are observed between the polymeric chains, which is in strong contrast to other 
hydrogen-rich materials that are stable against decomposition and exhibit a 3D network of strongly interacting 
host atoms and hydrogen~\cite{Disilane_JAFL,zhang_phase_2015}. With increasing hydrogen content the dimensionality of the polymers 
decreases from layers and sheets (PH, PH$_2$) to chains (PH$_3$, PH$_4$), until over-saturation is reached and H$_2$ molecules precipitate, 
intercalating the P-H polymers in PH$_5$ and PH$_6$.

Fig.~\ref{fig:enthalpy} shows that, although none of the binary phases are thermodynamically stable, some lie very close to the convex hull at
around 120\,GPa. Therefore, an in-depth investigation was conducted to assess whether effects going beyond those included in our MHM searches could stabilize any of 
these phases in the range between 100 and 150\,GPa. 

\begin{figure}[t]
\includegraphics[width=1.00\columnwidth,angle=0]{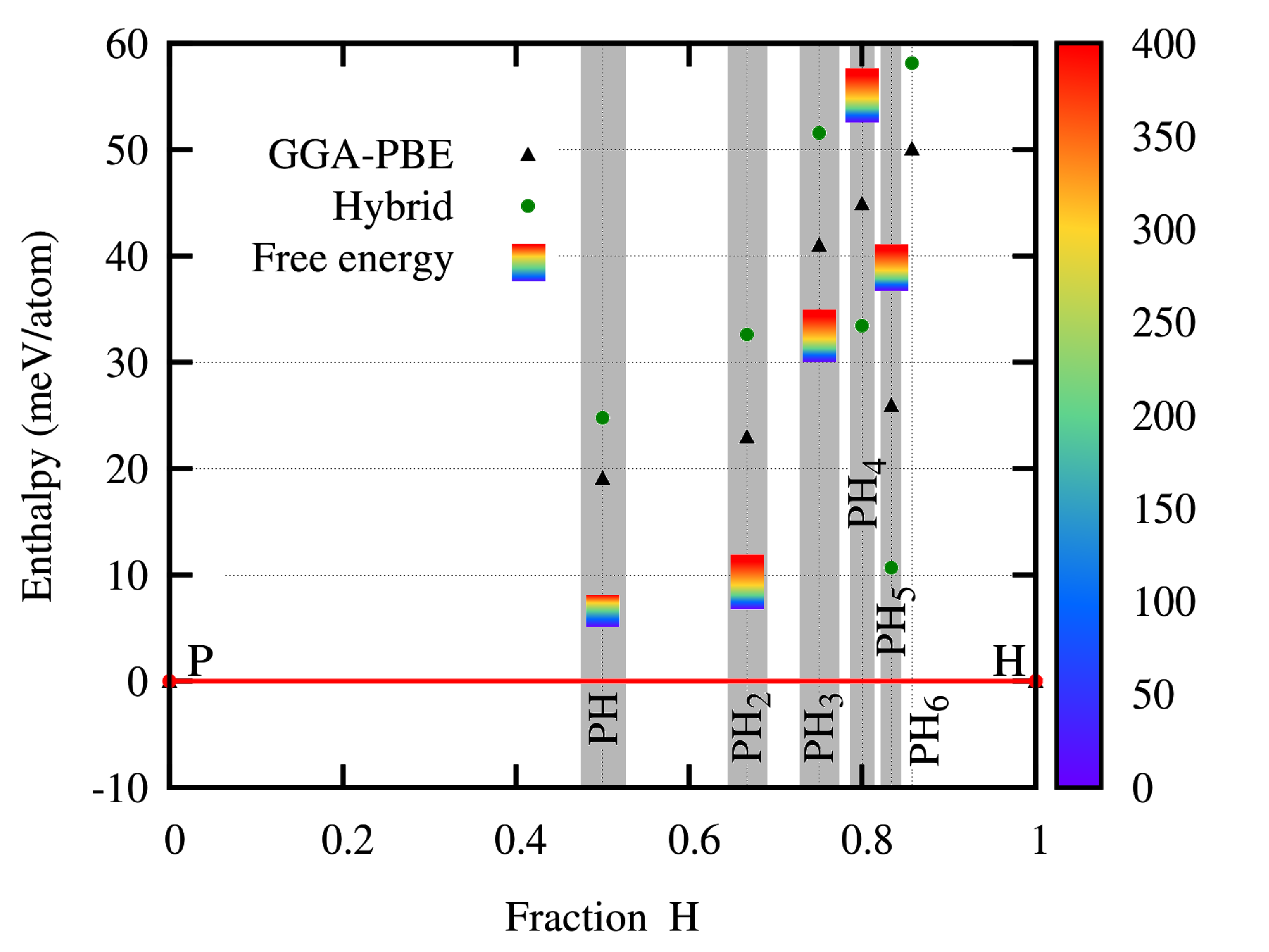}
\caption{(Color online) Predicted formation enthalpies of PH$_n$ with respect to decomposition into P and H at 120\,GPa. 
The solid red line denotes the convex hull of stability. 
Black triangles show PBE and green squares hybrid functional values (HSE06). 
The color-gradient scale indicate the free-energy within the harmonic approximation at temperatures up to 400\,K  
(colorbar in Kelvin on the right) as computed on top of PBE energies.}
 \label{fig:hull}
\end{figure}

A detailed analysis of the phase stability at 120\,GPa is shown in Fig.~\ref{fig:hull}. 
We first investigated how the use of different exchange-correlation functionals influences the stability of the various compounds. 
PBE values are shown as black triangles in Fig.~\ref{fig:hull}. 
In the local density approximation (not shown) we obtain fairly similar phase stabilities. 
We also report results obtained by means of the more demanding  Heyd-Scuseria-Ernzerhof (HSE) 
hybrid functional~\cite{paier_screened_2006,heyd_energy_2005,heyd_erratum:_2006,paier_erratum:_2006}.  
Although hybrids are considered to give better results than semilocal functionals for structural 
and electronic properties in semiconductors~\cite{heyd_energy_2005} and intermetallic 
alloys~\cite{zhang_nonlocal_2014}, they have not been systematically validated for metallic systems, 
where severe anomalies in the lattice stability and electronic 
properties have been recently reported~\cite{Gao_arXiv1504.06259}.
The HSE results, plotted as green squares in Fig.~\ref{fig:hull}, show nevertheless a similar trend as obtained with PBE and LDA. HSE predicts slightly higher formation enthalpies than PBE for PH, PH$_2$, PH$_3$ and PH$_6$, and lower values for PH$_4$ and PH$_5$. Finally, the semi-empirical DFT-D2 method of Grimme~\cite{grimme_2006}, which takes into account van-der-Waals interactions, consistently predicted a lower stability of the binary phases compared to PBE. The positive formation enthalpies predicted by all four exchange-correlation functionals strongly suggest that none of the binary PH$_n$ phases are in fact thermodynamically stable. 

Next, we investigated the influence of vibrational entropy and zero-point energy (ZPE) on the phase stability. 
Phonon calculations were carried out with the frozen phonon approach as implemented in the {\sc phonopy} 
package~\cite{phonopy} with sufficiently large super-cells. 
The ZPE and free energy within the harmonic approximation at temperatures of 100, 200, 300 and 400\,K are shown as colored gradients in Fig.~\ref{fig:hull}. 
There is a clear distinction in the vibrational effect on the stability for PH$_n$-phases 
with high and low hydrogen content: PH, PH$_2$ and PH$_3$ increase in stability, whereas PH$_4$ and PH$_5$ decrease. 
This behavior can be attributed to the existence of H$_2$ molecules in the crystal lattice of the hydrogen-rich phases which 
induce high-energy vibrational modes responsible for an increased ZPE.

For PH, formation energy is less than 5\,meV/atom at 0\,K, whereas for PH$_2$ and PH$_3$ it is about 7\,meV/atom and 30\,meV/atom, respectively. 
Comparing the free energies at finite temperatures reveals that all phases are destabilized by vibrational entropy by a few meV, 
pushing them away from the convex hull with increasing temperature. Since anharmonic effects become increasingly important at higher temperatures we recomputed 
the free energy for PH, PH$_2$ and PH$_3$ within the quasi-harmonic approximation, taking into account the thermal lattice expansion. 
These calculations, however, showed only minor effects on the stabilities and the distance from the convex hull 
of all three binary phases slightly increased by a few meV per atom. 

Having ruled out that any of the standard corrections to the electronic enthalpy can stabilize our 
phases in the experimentally relevant regime, we are left with the question of which phase and chemical composition 
was experimentally observed by Drozdov~\textit{et al.} We can rule out that the superconductivity observed is due to some 
residual elemental phase, as hydrogen is not metallic at these pressures and elemental phosphorus displays much lower critical
temperatures~\cite{Karuzawa_JoP_blackP_hipress160GPa_2002}, not higher than a few Kelvin.  

On the other hand, our phonon calculations show that the ground states at fixed compositions of PH$_1$, PH$_2$ and PH$_3$ are all dynamically stable, 
i.e. with no imaginary modes in the whole Brillouin zone. Therefore, one of these metastable phases might have been synthesized 
through a non-equilibrium process or possibly through anisotropic stress in the anvil cell, as the order of magnitude of the relevant 
energy barriers in these cases could be in the range of few tens of meV. In fact, the compositions closest to the hull are PH and PH$_2$, indicating a 
possible decomposition of the original phosphine molecules (PH$_3$) under pressure into PH and PH$_2$, similarly to what is observed in sulfur-hydrides where the initial, molecular SH$_2$ decomposes under pressure to form SH$_3$.
 
To assess whether any of these three metastable phases could be a reasonable candidate for superconductivity, 
we calculated the Eliashberg spectral functions for the electron-phonon ($ep$) interaction:
\begin{equation}
 \alpha^2 F(\omega) = \frac{1}{N_{E_F}} \sum \limits_{\mathbf{k} \mathbf{q},\nu} |g_{\mathbf{k},\mathbf{k}+\mathbf{q},\nu}|^2 \delta(\epsilon_\mathbf{k}) \delta(\epsilon_{\mathbf{k}+\mathbf{q}}) \delta(\omega-\omega_{\mathbf{q},\nu})~, \label{eq:a2F}
\end{equation}
where $N_{E_F}$ is the DOS at the Fermi level, $\omega_{\mathbf{q},\nu}$ is the phonon frequency of mode $\nu$ at wavevector $\mathbf{q}$ and 
$|g_{\mathbf{k},\mathbf{k}+\mathbf{q},\nu}|$ is the electron-phonon matrix element between two electronic states with momenta $\mathbf{k}$ and $\mathbf{k+q}$ at the Fermi level.

From the Eliashberg function we also obtained the electron-phonon coupling constant $\lambda$ and the logarithmic average phonon frequency 
\omlog~\cite{Carbotte_RMP1990,AllenMitrovic1983} (which, in the McMillan-Allen-Dynes parametrization for \tc, sets the energy scale for the phononic pairing). 
In order to avoid using the empirical $\mu^*$ parameter and bias the theoretical predictions towards the experiments, we compute the critical temperatures within density functional theory for superconductors (SCDFT), 
which uses as input the Eliashberg function in an isotropic approximation, while the 
residual Coulomb forces in the Cooper pairing are included within the static 
random phase approximation~\cite{OGK_SCDFT_PRL1988,Lueders_SCDFT_PRB2005,Marques_SCDFT_PRB2005,Massidda_SUST_CoulombSCDFT_2009}, 
as used in Ref.~\onlinecite{FloresSanna_H3Se_Arxiv2015,Flores-Sanna_PRBhoneycombs}.

\begin{figure}[t]
\includegraphics[width=1.0\columnwidth,angle=0]{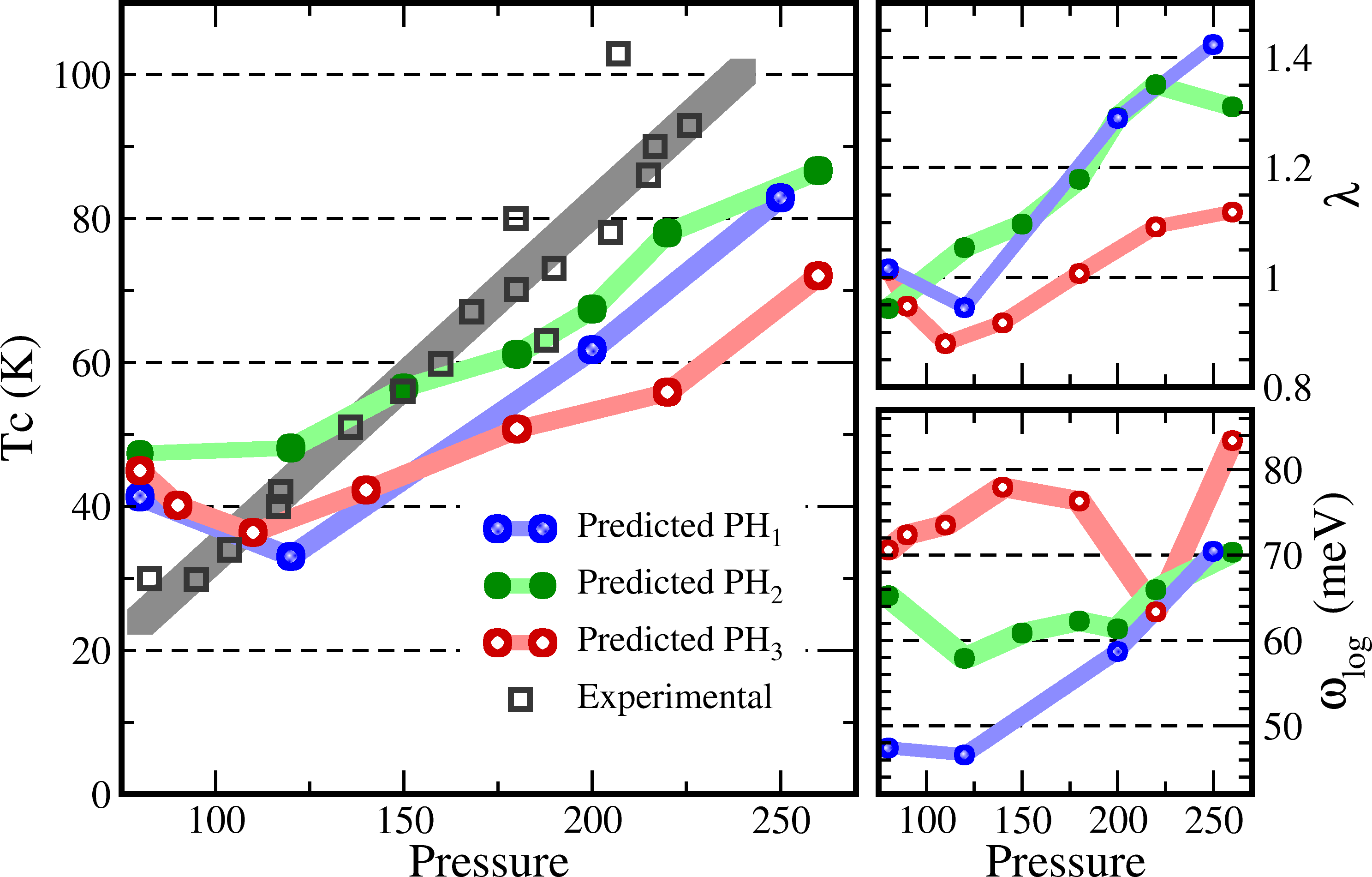}
  \caption{(Color online) Left:  SCDFT calculated critical temperatures \tc\,for PH (blue), PH$_2$ (green) and PH$_3$ (red) as a function of pressure. 
  Experimental \tc\ by resistivity measurements from Drozdov~\textit{et al.}~\cite{Drozdov_ph3_arxiv2015} are shown in black squares.  
  Overall a good agreement is found between the experimental values and those for PH, PH$_2$ and PH$_3$. 
  Although the PH$_2$  composition shows significantly better agreement. 
  Right: trend in pressure of the (BCS-like) electron phonon coupling coefficient $\lambda$ (top panel) and of the phononic characteristic frequency \omlog\ (bottom panel) as a function of pressure. }
 \label{fig:tc_values}
\end{figure}

The values of \tc \ are shown in Fig.~\ref{fig:tc_values} and are compared to the experimental values 
reported by Drozdov~\textit{et al.}~\cite{Drozdov_ph3_arxiv2015}. 
In the two right panels of the same figure we show $\lambda$ and \omlog\ for the three phases as a function of pressure
~\footnote{Electronic bands, vibrational frequencies and electron-phonon matrix elements were computed within density functional 
perturbation theory,~\cite{DFPT_S.Baroni,2n+1_Gonzepaper,Savrasov2_PRB} as implemented in the {\sc Quantum Espresso} and  {\sc abinit} code. 
We employed norm-conserving pseudopotentials for H and P, with a plane-wave cut-off of 80\,Ry; phonon frequencies and electron-phonon were 
computed on coarse (8$^3$, 6$^3$ and 4$^3$\,k points for PH, PH$_2$ and PH$_3$ respectively), 
and interpolated on denser grids using force constants; for the electronic integration, we employed Monkhorst-Pack grids 
of 8$^3$\,$\mathbf{k}$ points for the self-consistent calculations; much denser (up to 30$^3$ points) 
grids were used for Fermi-surface integrations.}.

Our calculated behavior of \tc\ with respect to pressure shows a fair agreement with experiments for all three structures (PH, PH$_2$ and PH$_3$); 
The best agreement is found for PH$_2$, which has a \tc \ of 40\,K at 100\,GPa that increases under pressure and 
reaches a maximum value of 78\,K at 220\,GPa. 
The PH system shows the best agreement in the rate at which \tc\ grows with pressure in the 120-260\,GPa window 
($dTc/dP\simeq 0.4 (K/GPa)$), while showing a high pressure shift of about 20\,GPa with respect to the experimental data.  

\begin{figure}[t]
\includegraphics[width=1.0\columnwidth,angle=0]{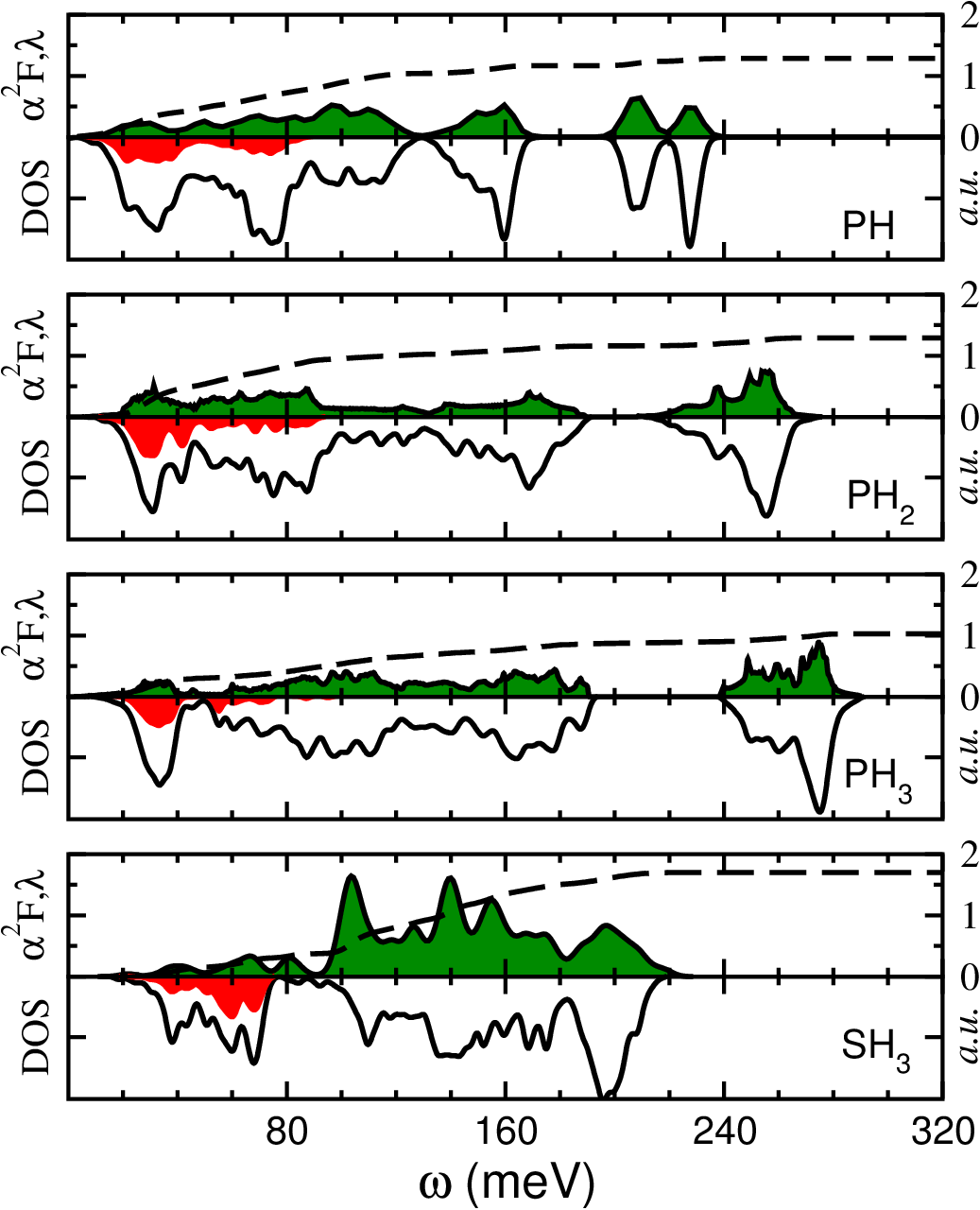}
  \caption{(Color online)
 Eliashberg spectral function and frequency-dependent $ep$ coupling parameters $\lambda(\omega)$ -- top panels -- 
 and phonon density of states (DOS) for the three PH$_n$ compounds considered in the present study and SH$_3$ at 200\,GPa. 
 The shaded area (red) in the DOS panels is the P/S partial phonon DOS.}
 \label{fig:alpha}
\end{figure}

The similar superconducting behavior of the three compounds results from the compensation of different behaviors
of $\lambda$ and \omlog, shown in the right panels of the same figure. These reflect 
different features of the three Eliashberg functions, shown in Fig.~\ref{fig:alpha}, together with that of SH$_3$ for reference. 
The PH$_{1,2,3}$ spectra have an overall
similar shape, {\em i.e.} they are roughly proportional to the phonon density of states,
dominated by P modes at low frequencies (< 80\,meV), and hydrogen modes at high frequencies. 

A gap separates the hydrogen bond-bending vibrations 
from the rest of the spectrum; this part moves to higher energies with increasing hydrogen content, but has
very little influence on \tc\ due to the high frequencies involved. 
The spectrum of SH$_3$ is more compact, extending up to 200\,meV, having peaks of the $\alpha^2 F(\omega)$ higher, 
this result in a large $\lambda (\simeq 1.9)$. 

As more and more theoretical predictions of new superconducting hydrides are available, 
there is an increasing effort towards a systematic understanding of the factors leading
to high-\tc~\cite{zhang_phase_2015,kim_general_2010}. 
It is becoming clear that the original idea of Ashcroft~\cite{Ashcroft_PRL2004}, 
that the heavier atoms merely exert chemical pressure on the hydrogen lattice, 
is oversimplified, and other factors, such as the formation of strong atomic bonds 
between H and the heavier elements, play a crucial role.
A good indicator of the tendency of a material to follow one or another behavior is the electronegativity of the heavy atom. 
Atoms which are less electronegative than H tend to form solids which contain H$_2$ units with large characteristic vibration frequencies, 
but relatively weak matrix elements (electron-phonon coupling). 
More electronegative atoms, on the other hand, tend to form polar-covalent bonds, 
which couple strongly to phonons~\cite{Heil-Boeri_PRB2015,zhang_phase_2015}. 

Phosphorus, with an electronegativity equal to that of hydrogen, lies on the 
border between the two regions and for 
the three hydrides considered in this study the electronic structure indicates a strong P-H hybridization 
in the whole energy range (see supplemental material). The values of the electronic density of state (DOS) 
at the Fermi level -- $N_{E_F}$ -- are comparable ($0.31-0.38\,st/eV\,f.u.$), and sensibly lower than those in SH$_3$ (0.54).
The ratio $\eta=\lambda/N_{E_F}$, which can give an indication of the ``stiffness'' of the underlying lattice, 
ranges from 3 in PH$_3$ to 3.8 in PH, and is 3.6 in SH$_3$.

In practice,  PH$_{1,2,3}$ and SH$_3$ have a quite similar
bonding and lattice stiffness to each other:
what makes the latter exhibit a yet unmatched record \tc,\ is the extremely high value of its $N_{E_F}$, due to
the presence of the van-Hove singularity close to the Fermi level~\cite{quan_impact_2015,ortenzi_TB_2015}. 
As the Fermi level of PH$_2$ and PH$_3$ falls in a shallow minimum of the density of states (see supplemental material), 
the same effect could be used to increase their \tc, by doping a small fraction of S or Si impurities into the samples. 

In conclusion, the phase stability and superconducting properties of the recently reported PH$_3$ compound under pressure have been investigated 
with \textit{ab initio} calculations. 
Our extensive structural searches show that the hydrogen-rich phosphorus phases are thermodynamically unstable in the high-pressure regime of the phase diagram. 
Although including vibrational zero-point effects in our calculations improves the stability of PH, PH$_2$ and PH$_3$, they remain metastable. 
Nevertheless, in view of their small distance from the convex hull, any of these structures may have been synthesized through non-equilibrium effects,
or by anharmonic effects, as discussed in \cite{Errea_anhaPRL2015,Engel_anharmonic_PRX2015}.
Several of them were predicted to be good phonon-mediated superconductors and could thus in principle account for the measured high-\tc\, in experiments. 
In our opinion, the phase/composition that yields the best agreement with experiments are 
PH and PH$_2$, which are both close to the convex hull ($<$ 7\,meV/atom) 
and would show a pressure dependence in \tc\ similar to the experimental measurements.  
While experimentally it is still unclear if the observed high-\tc \ can be fully attributed to a 
phase of a binary phosphorus-hydride, and further experiments are called for to validate both our predictions and the results reported 
by Drozdov~\textit{et al.}~\cite{Drozdov_ph3_arxiv2015}.
Finally, our calculations give a strong indication that the observed critical temperatures are 
those to be expected for the low energy, low-H content phases of P-H systems.  

\underline{Note:} While writing the present manuscript, 
we became aware of two works which report structural searches of phosphorus-hydrides, 
employing evolutionary algorithms~\cite{shamp_decomposition_2015,zhang_phase_2015}. 
Both studies agree with our conclusions that none of the predicted phases is thermodynamically stable in the pressure range of 
100-200\,GPa. {\em Shamp et al.} also suggest an $I4/mmm$ PH$_2$ structure as a possible candidate for the high-\tc \ phase. 
They also report a \textit{polymeric} PH$_2$ structure which is more stable at low pressures. 
Based on our calculations, however, the \textit{polymeric} structures for PH$_2$ show higher enthalpies than the $I4/mmm$ structure.

\begin{acknowledgments}
J.A.F.-L. acknowledges the EU's 7th  Framework Marie-Curie Program within the ``ExMaMa'' Project (329386). 
M.A. gratefully acknowledges support from the Novartis Universit\"{a}t Basel Excellence Scholarship for Life Sciences and the Swiss National Science Foundation. 
C.H. and L.B. acknowledge funding from the FWF-SFB ViCoM F41 P15 and computational resources from the dCluster of the Graz University of Technology 
and the VSC3 of the Vienna University of Technology. 
CW acknowledges support by the U.S. Department of Energy, Office of Science, Basic Energy Sciences, under Grant DEFG02-07ER46433. 
Computational resources from the Swiss National Supercomputing Center (CSCS) in Lugano (project s499) and the National 
Energy Research Scientific Computing Center, which is supported by the Office of Science of the U.S. Department of Energy 
under Contract No. DE-AC02-05CH11231, are acknowledged. This work was done within the NCCR MARVEL project.

\end{acknowledgments}
\bibliographystyle{apsrev4-1}
\bibliography{paper}
\end{document}